\documentclass[12pt]{article}
\begin{document}

\title{The Confining Interaction and Radiative Decays of Heavy
Quarkonia}

\author{T.A. L\"ahde\footnote{talahde@rock.helsinki.fi},
 C.J. Nyf\"alt\footnote{nyfalt@rock.helsinki.fi} and
D.O. Riska\footnote{riska@rock.helsinki.fi}}
\date{}
\maketitle

\centerline{\it Department of Physics}
\centerline{\it POB 9, 00014 University of Helsinki}
\centerline{\it Finland}

\vspace{1cm}

\begin{abstract}
The radiative spin-flip transition rates of heavy quarkonium 
states depend
sensitively on the matrix elements of the effective confining
interaction through the associated two-quark exchange current
operator. 
The Hamiltonian model based on a scalar linear confining 
interaction with 
a single 
gluon exchange 
hyperfine term is shown to provide an adequate
description of the $J/\psi\rightarrow \eta_c\gamma$ and
$\psi(2S)\rightarrow \eta_c\gamma$ decay widths 
once the relativistic
single quark magnetic moment operator 
is treated without approximation
and the exchange current is taken into account.
Predictions are given for the radiative spin-flip decay widths of
the $1S$,$2S$ and $3S$ states of the $c\bar c$, $b\bar b$ and 
$B_c^+$ systems. 
In the $B_c^+$ system the gluon exchange
current also contributes to the spin-flip transition rates.

\end{abstract}

\newpage

\section{Introduction}

The spectra of the $c\bar c$ and $b\bar b$ meson systems are fairly
well reproduced by a Schr\"odinger equation based description with an
interaction Hamiltonian formed of a scalar linear confining
interaction and a single gluon exchange model for the hyperfine
interaction \cite{Godfrey,Quigg}. This model is qualitatively
supported by numerical construction of the effective interaction by
lattice methods \cite{Bali1,Bali2}. For this linear
confinement + gluon exchange model the parameter freedom has recently
been narrowed considerably by numerical precision determination of the
effective quark-gluon coupling strength $\alpha_s$ 
\cite{Lepage1,Lepage}.\\

\begin{figure}[ht!]
\begin{center}
\begin{picture} (370,220)

\put(80,20){\line(0,1){100}}
\put(150,20){\line(0,1){180}}
\put(80,120){\line(-1,-1){40}}
\put(40,80){\line(0,1){120}}

\multiput(82.5,120)(10,0){7}{\line(1,0){5}}
\multiput(-2.5,80)(10,0){5}{\oval(5,5)[t]}
\multiput(-7.5,80)(10,0){5}{\oval(5,5)[b]}

\put(250,20){\line(0,1){100}}
\put(360,20){\line(0,1){180}}
\put(250,120){\line(1,-1){40}}
\put(290,80){\line(0,1){120}}

\multiput(292.5,80)(10,0){7}{\line(1,0){5}}
\multiput(207.5,120)(10,0){5}{\oval(5,5)[t]}
\multiput(202.5,120)(10,0){5}{\oval(5,5)[b]}

\end{picture}
\end{center}
\caption{Exchange current operators associated with the effective
scalar confining and gluon exchange interactions, with
intermediate virtual $q\bar q$ excitations.} \label{fey}
\end{figure}
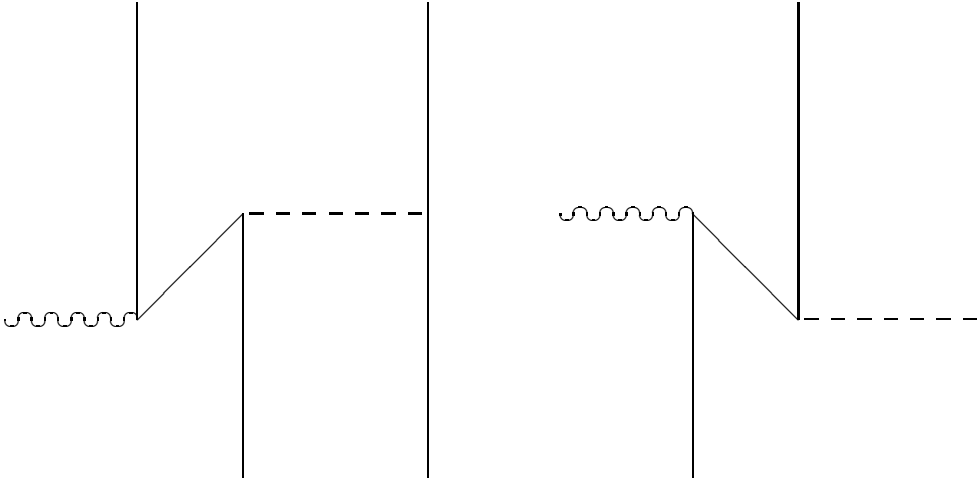

While the spectra of the $c\bar c$ and $b\bar b$ systems are 
fairly well accounted for by the simple linear confinement + gluon
exchange interaction model, the situation concerning the radiative
widths of the spin-flip transitions $J/\psi\rightarrow \eta_c\gamma$
and $\psi(2S)\rightarrow \eta_c\gamma$ has remained unsettled
\cite{Grotch1,Grotch2,Snellman}. We show here that these 
radiative transitions
strengths are in fact also satisfactorily described
by the conventional quantum mechanical
framework, provided that the full Dirac structure of the quark current
operators is taken into account along with the exchange current
operator that is 
generated by coupling to intermediate negative energy states
by the linear scalar confining operator Fig.\ref{fey}, 
once the same Hamiltonian 
is used to
generate both the wave functions and the current operators.
This exchange current contribution is decisive for
achieving agreement between the calculated and
measured M1 transition rates of the $J/\psi$ and the $\psi'$. 
The exchange current operator is required by current 
conservation with the confining interaction.\\

The contribution of the exchange current operator that is 
associated with the linear potential is
essential for reproduction of the empirical decay widths. Moreover
this operator only arises under the assumption that the effective
confining interaction is a Lorentz scalar (``$S$''), which 
supports the
consistency requirement of ref.\cite{Gromes}. If the Lorentz
invariant structure of the confining interaction were a vector
invariant (``$V$''), the associated exchange current operator would
contain no spin-flip term when the quark and antiquarks
have equal mass as in the $c\bar c$ system, 
and agreement with the empirical decay
widths would be excluded. \\

\newpage

In the case of the $B_c^+$ system, the first state of which
has recently been observed \cite{Fermi}, the exchange current that is 
associated with the vector gluon exchange
interaction in contrast also contains a component that is 
antisymmetric in the spins of the
quarks and antiquarks, which in this case have 
unequal mass. That exchange
current consequently also plays a role in the
spin-flip transitions of the $B_c^+$ system, because of the
large difference in mass between the $c$ and $b$ quarks.
The role of these exchange currents is
shown here by a calculation of the
the decay widths for the radiative spin-flip 
transitions $B_c^+(J=1,\,
nS)\rightarrow B_c^+(J=0,\, 1S)\gamma$ $(n=1,2,3)$. The decay
widths are calculated here with full account of both the confinement
and gluon exchange currents. 
The radiative decay widths have been calculated in the
non-relativistic impulse approximation without
consideration of the exchange current contributions in
ref.\cite{Quigg1}.
Experimental determination of these decay
widths should be decisive for settling the quantitative
importance of the exchange current operators and
the validity of the present
phenomenological approach to the structure of heavy quarkonia.\\

The $B_c^+$ systems differ from the $c\bar c$ and $b\bar b$ systems
also in that the $B_c^+$ meson have magnetic moments. These also 
obtain considerable exchange current contributions which are 
calculated here.\\

The framework used here is similar to that employed
in refs. \cite{Grotch1,Grotch2} for the calculation
of the rates of the M1 transitions in the $c\bar c$
and $b\bar b$ systems. The affirmative conclusion about
the significance of the exchange current contributions
here is new. The numerical values differ from those
in refs.\cite {Grotch1,Grotch2} partly because the unapproximated
Dirac magnetic moment operator is considered here, and
because of an overestimate of the lowest order relativistic
correction to the quark magnetic moment operator.\\

This paper is divided into 5 sections. In section \ref{mh-sec} we describe
the
Hamiltonian model for the heavy quarkonium systems, and demonstrate
that it describes the spectra of the $c\bar c$ and $b\bar b$ mesons
satisfactorily. In section \ref{bc-sec} we show the predicted spectrum of
the
$B_c^+$ system. In section \ref{curr-sec} we describe the model for the
current
operator for the heavy quarkonium system. The corresponding calculated
decay widths for the radiative spin-flip transitions of the $c\bar c$,
$b\bar b$ and $B_c^+$ mesons are presented in section \ref{tran-sec}.
Finally
section \ref{disc-sec}  contains a concluding discussion.\\

\section{Model Hamiltonian} \label{mh-sec}

\subsection{Hamiltonian}

The linear scalar confinement + gluon exchange hyperfine interaction
Hamiltonian model for heavy quarkonia $Q\bar q$, is formed of quarks
$Q$ and antiquarks $\bar q$ with in general unequal mass 
\begin{equation}
H=H_{kin}+V_C+V_G \label{ham}
\end{equation}
where $H_{kin}$ is the kinetic energy term and $V_C$ and $V_G$ are the
potentials that describe the confining and gluon exchange interactions
respectively. \\

The kinetic energy term, which in heavy quark effective theory is
taken to include the terms to order $m^{-4}$ is 

\begin{equation}
H_{kin}=m_Q+m_{\bar q}+\frac{p^2}{2m_r}-\frac{1}{8}\left(
\frac{m_Q^3+m_{\bar q}^3}{m_Q^3m_{\bar q}^3}\right)p^4,
\label{kinerg}
\end{equation}
where $m_r$ is the reduced mass $m_r=m_Qm_{\bar q}/(m_Q+m_{\bar q})$
and $\vec p$ is the relative momentum. The term of $\mathcal{O}(p^4)$, which
will be treated as a perturbation increases in importance with
excitation number, and is non-negligible because of the compact spatial
extent of the confined quarkonium wave functions. \\

The scalar confining potential has the form (to order
$m^{-2}$):
\begin{eqnarray}
&&V_C(r)=cr\left\{1-\frac{\vec p^2}{2}\frac{m_Q^2+m_{\bar q}^2}
{m_Q^2m_{\bar q}^2}\right\}
-\frac{c}{4r} \frac{m_Q^2+m_{\bar q}^2}
{m_Q^2m_{\bar q}^2} \vec S\cdot
\vec L \cr\cr
&&+\frac{c}{8r}\frac{m_Q^2-m_{\bar q}^2}
{m_Q^2m_{\bar q}^2}(\vec{\sigma}_Q
-\vec{\sigma}_q)\cdot \vec{L}.
\label{conf}
\end{eqnarray}
The last antisymmetric spin-orbit term in this interaction vanishes
for equal mass quarkonia. \\

The single gluon exchange interaction potential to order $m^{-2}$ has
the form
\begin{eqnarray}
&&V_G(r)=-\frac{4}{3} \alpha_s\left\{ \frac{1}{r}
-\pi \delta^{(3)}(\vec r)
\frac{m_Q^2+m_{\bar q}^2}{2m_Q^2 m_{\bar q}^2} 
+\frac{ \vec p^2} {r m_Q m_{\bar q} }\right\}\cr\cr
&&+\frac{8\pi}{9}\frac{\alpha_s}{m_{Q}m_{\bar{q}}}
\delta^{(3)}(\vec r)\vec{\sigma}^1\cdot\vec{\sigma}^2
+\frac{\alpha_s}{3m_Qm_{\bar q}}\frac{1}{r^3}S_{12}\cr\cr
&&+\frac{2\alpha_s}{3r^{3}}\left\{\frac{m_Q^2+m_{\bar q}^2}{2m_Q^2m_{\bar
q}^2}+\frac{2}{m_Qm_{\bar q}}\right\}\vec S\cdot \vec L
+\frac{\alpha_s}{6r^3}\frac{m_Q^2-m_{\bar q}^2}{m_Q^2m_{\bar q}^2}
(\vec{\sigma}_Q-\vec{\sigma}_{\bar q}) \cdot \vec L. \label{exch}
\end{eqnarray}
Here $S_{12}\equiv 3\vec{\sigma}_{Q}\cdot \hat r\vec{\sigma}_{\bar q}
\cdot \hat r-\vec{\sigma}_Q\cdot \vec{\sigma}_{\bar q}$ is the usual
tensor interaction operator.\\

The interaction operator constructed to order $m^{-2}$ by lattice
methods in refs. \cite{Bali1,Bali2} contains in addition
to the operators of the form
contained in the model Hamiltonian (\ref{ham}) that involve the non-local
operator $\vec p/m$ to second order, but which do 
not significantly affect
the calculated spectra. 
We do not include such here, mainly because they lack an
immediate dynamical interpretation.
The interaction operator derived in refs.
\cite{Bali1,Bali2} has a string tension $c$ of $\sim 1.1$ GeV/fm,
with a slight flavor dependence.\\

\begin{table}[h]
\noindent
\begin{center}
\begin{tabular}{c|c|c|c|}
 & $\mathbf{c\bar c}$ & $\mathbf{B_c^+}$ & $\mathbf{b \bar b}$\\ \hline
$\mathbf{c}$ & $1.11$ GeV/fm & $1.11$ GeV/fm & $1.11$ GeV/fm\\ \hline
$\mathbf{\alpha_s}$ & 0.40 & 0.32 & 0.29\\ \hline
$\mathbf{m_c}$ & $1.48$ GeV & $1.48$ GeV & --\\ \hline
$\mathbf{m_b}$ & -- & $4.72$ GeV & $4.72 $ GeV\\ \hline
\end{tabular}
\end{center} 
\caption{Parameter values for the model Hamiltonian (\ref{ham})
used here.} \label{para}
\end{table}

We shall here use parameter values close to those suggested in
\cite{Bali1,Bali2,Lepage}, but determine the precise values by
phenomenological fits to the $c\bar c$ and $b\bar b$ spectra. The
values used are listed in Table \ref{para}. 
The values for the color fine structure constant 
$\alpha_s$ in the table
are close to those obtained by lattice methods
in refs. \cite{Lepage1,Lepage}: $\alpha_s=0.38$ at the charmonium scale  
$m_c\sim 1.3$ and 
$\alpha_s=0.22$ at the bottom scale $m_b\sim 4.1$ GeV
respectively. 
These values agree well with the inverse
logarithmic momentum dependence predicted by perturbative $QCD$.
The phenomenologically
determined quark masses in Table \ref{para} should
be viewed as constituent quark masses. 
The overall
description of the empirical spectra corresponds to that
achieved in refs. \cite{Bali1,Bali2}.\\

\subsection{The $c\bar c$ and $b\bar b$ Spectra}

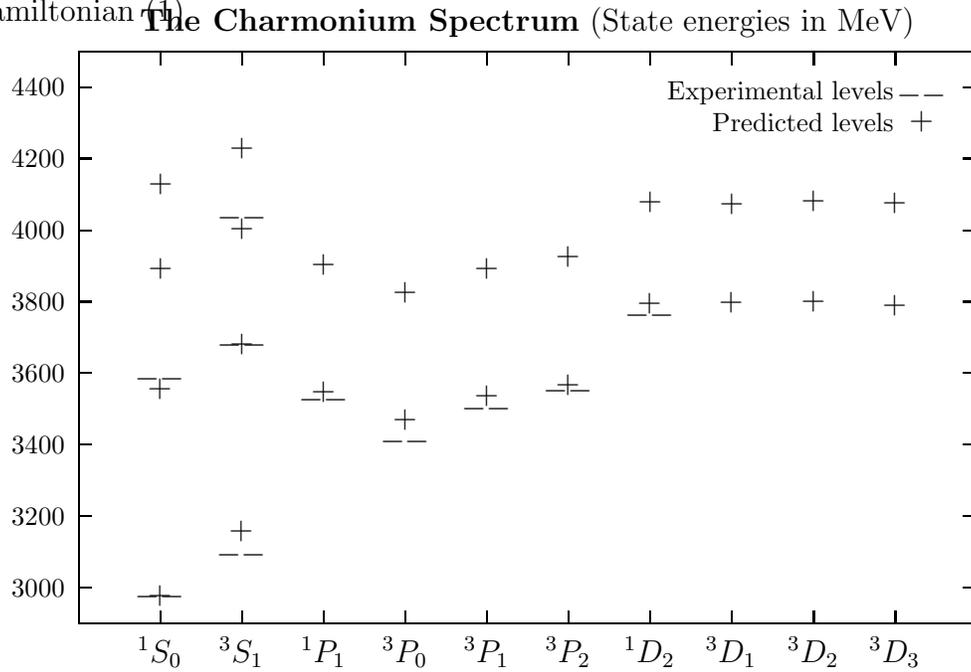
\begin{figure}[hb!]
\caption{Charmonium spectrum with $m_{c}=1480$ MeV
obtained from the Hamiltonian (1)} \label{cc}
\begin{center}
\setlength{\unitlength}{0.240900pt}
\ifx\plotpoint\undefined\newsavebox{\plotpoint}\fi
\sbox{\plotpoint}{\rule[-0.200pt]{0.400pt}{0.400pt}}%
\begin{picture}(1650,500)(0,450)
\font\gnuplot=cmr10 at 10pt
\gnuplot
\sbox{\plotpoint}{\rule[-0.200pt]{0.400pt}{0.400pt}}%
\put(176.0,124.0){\rule[-0.200pt]{4.818pt}{0.400pt}}
\put(154,124){\makebox(0,0)[r]{3000}}
\put(1566.0,124.0){\rule[-0.200pt]{4.818pt}{0.400pt}}
\put(176.0,236.0){\rule[-0.200pt]{4.818pt}{0.400pt}}
\put(154,236){\makebox(0,0)[r]{3200}}
\put(1566.0,236.0){\rule[-0.200pt]{4.818pt}{0.400pt}}
\put(176.0,349.0){\rule[-0.200pt]{4.818pt}{0.400pt}}
\put(154,349){\makebox(0,0)[r]{3400}}
\put(1566.0,349.0){\rule[-0.200pt]{4.818pt}{0.400pt}}
\put(176.0,461.0){\rule[-0.200pt]{4.818pt}{0.400pt}}
\put(154,461){\makebox(0,0)[r]{3600}}
\put(1566.0,461.0){\rule[-0.200pt]{4.818pt}{0.400pt}}
\put(176.0,573.0){\rule[-0.200pt]{4.818pt}{0.400pt}}
\put(154,573){\makebox(0,0)[r]{3800}}
\put(1566.0,573.0){\rule[-0.200pt]{4.818pt}{0.400pt}}
\put(176.0,685.0){\rule[-0.200pt]{4.818pt}{0.400pt}}
\put(154,685){\makebox(0,0)[r]{4000}}
\put(1566.0,685.0){\rule[-0.200pt]{4.818pt}{0.400pt}}
\put(176.0,798.0){\rule[-0.200pt]{4.818pt}{0.400pt}}
\put(154,798){\makebox(0,0)[r]{4200}}
\put(1566.0,798.0){\rule[-0.200pt]{4.818pt}{0.400pt}}
\put(176.0,910.0){\rule[-0.200pt]{4.818pt}{0.400pt}}
\put(154,910){\makebox(0,0)[r]{4400}}
\put(1566.0,910.0){\rule[-0.200pt]{4.818pt}{0.400pt}}
\put(304.0,68.0){\rule[-0.200pt]{0.400pt}{4.818pt}}
\put(304,23){\makebox(0,0){$ ^{1}S_{0}$}}
\put(304.0,946.0){\rule[-0.200pt]{0.400pt}{4.818pt}}
\put(432.0,68.0){\rule[-0.200pt]{0.400pt}{4.818pt}}
\put(432,23){\makebox(0,0){$ ^{3}S_{1}$}}
\put(432.0,946.0){\rule[-0.200pt]{0.400pt}{4.818pt}}
\put(561.0,68.0){\rule[-0.200pt]{0.400pt}{4.818pt}}
\put(561,23){\makebox(0,0){$ ^{1}P_{1}$}}
\put(561.0,946.0){\rule[-0.200pt]{0.400pt}{4.818pt}}
\put(689.0,68.0){\rule[-0.200pt]{0.400pt}{4.818pt}}
\put(689,23){\makebox(0,0){$ ^{3}P_{0}$}}
\put(689.0,946.0){\rule[-0.200pt]{0.400pt}{4.818pt}}
\put(817.0,68.0){\rule[-0.200pt]{0.400pt}{4.818pt}}
\put(817,23){\makebox(0,0){$ ^{3}P_{1}$}}
\put(817.0,946.0){\rule[-0.200pt]{0.400pt}{4.818pt}}
\put(945.0,68.0){\rule[-0.200pt]{0.400pt}{4.818pt}}
\put(945,23){\makebox(0,0){$ ^{3}P_{2}$}}
\put(945.0,946.0){\rule[-0.200pt]{0.400pt}{4.818pt}}
\put(1073.0,68.0){\rule[-0.200pt]{0.400pt}{4.818pt}}
\put(1073,23){\makebox(0,0){$ ^{1}D_{2}$}}
\put(1073.0,946.0){\rule[-0.200pt]{0.400pt}{4.818pt}}
\put(1201.0,68.0){\rule[-0.200pt]{0.400pt}{4.818pt}}
\put(1201,23){\makebox(0,0){$ ^{3}D_{1}$}}
\put(1201.0,946.0){\rule[-0.200pt]{0.400pt}{4.818pt}}
\put(1330.0,68.0){\rule[-0.200pt]{0.400pt}{4.818pt}}
\put(1330,23){\makebox(0,0){$ ^{3}D_{2}$}}
\put(1330.0,946.0){\rule[-0.200pt]{0.400pt}{4.818pt}}
\put(1458.0,68.0){\rule[-0.200pt]{0.400pt}{4.818pt}}
\put(1458,23){\makebox(0,0){$ ^{3}D_{3}$}}
\put(1458.0,946.0){\rule[-0.200pt]{0.400pt}{4.818pt}}
\put(176.0,68.0){\rule[-0.200pt]{339.669pt}{0.400pt}}
\put(1586.0,68.0){\rule[-0.200pt]{0.400pt}{216.328pt}}
\put(176.0,966.0){\rule[-0.200pt]{339.669pt}{0.400pt}}
\put(881,1011){\makebox(0,0){\bf{The Charmonium Spectrum}\textmd{ (State
energies in MeV)}}}
\put(176.0,68.0){\rule[-0.200pt]{0.400pt}{216.328pt}}
\put(1456,901){\makebox(0,0)[r]{Experimental levels}}
\put(1500,901){\raisebox{-.8pt}{\makebox(0,0){$--$}}}
\put(304,113){\raisebox{-.8pt}{\makebox(0,0){$--$}}}
\put(304,455){\raisebox{-.8pt}{\makebox(0,0){$--$}}}
\put(432,179){\raisebox{-.8pt}{\makebox(0,0){$--$}}}
\put(433,509){\raisebox{-.8pt}{\makebox(0,0){$--$}}}
\put(433,708){\raisebox{-.8pt}{\makebox(0,0){$--$}}}
\put(561,422){\raisebox{-.8pt}{\makebox(0,0){$--$}}}
\put(689,357){\raisebox{-.8pt}{\makebox(0,0){$--$}}}
\put(817,409){\raisebox{-.8pt}{\makebox(0,0){$--$}}}
\put(945,436){\raisebox{-.8pt}{\makebox(0,0){$--$}}}
\put(1073,556){\raisebox{-.8pt}{\makebox(0,0){$--$}}}
\put(1456,856){\makebox(0,0)[r]{Predicted levels}}
\put(1500,856){\makebox(0,0){$+$}}
\put(304,112){\makebox(0,0){$+$}}
\put(304,437){\makebox(0,0){$+$}}
\put(305,626){\makebox(0,0){$+$}}
\put(305,758){\makebox(0,0){$+$}}
\put(432,213){\makebox(0,0){$+$}}
\put(433,507){\makebox(0,0){$+$}}
\put(433,688){\makebox(0,0){$+$}}
\put(433,815){\makebox(0,0){$+$}}
\put(561,431){\makebox(0,0){$+$}}
\put(561,631){\makebox(0,0){$+$}}
\put(689,388){\makebox(0,0){$+$}}
\put(689,588){\makebox(0,0){$+$}}
\put(817,426){\makebox(0,0){$+$}}
\put(817,625){\makebox(0,0){$+$}}
\put(945,443){\makebox(0,0){$+$}}
\put(945,644){\makebox(0,0){$+$}}
\put(1073,570){\makebox(0,0){$+$}}
\put(1074,730){\makebox(0,0){$+$}}
\put(1201,572){\makebox(0,0){$+$}}
\put(1202,727){\makebox(0,0){$+$}}
\put(1330,574){\makebox(0,0){$+$}}
\put(1330,732){\makebox(0,0){$+$}}
\put(1458,568){\makebox(0,0){$+$}}
\put(1458,729){\makebox(0,0){$+$}}
\end{picture}

\end{center}
\end{figure}

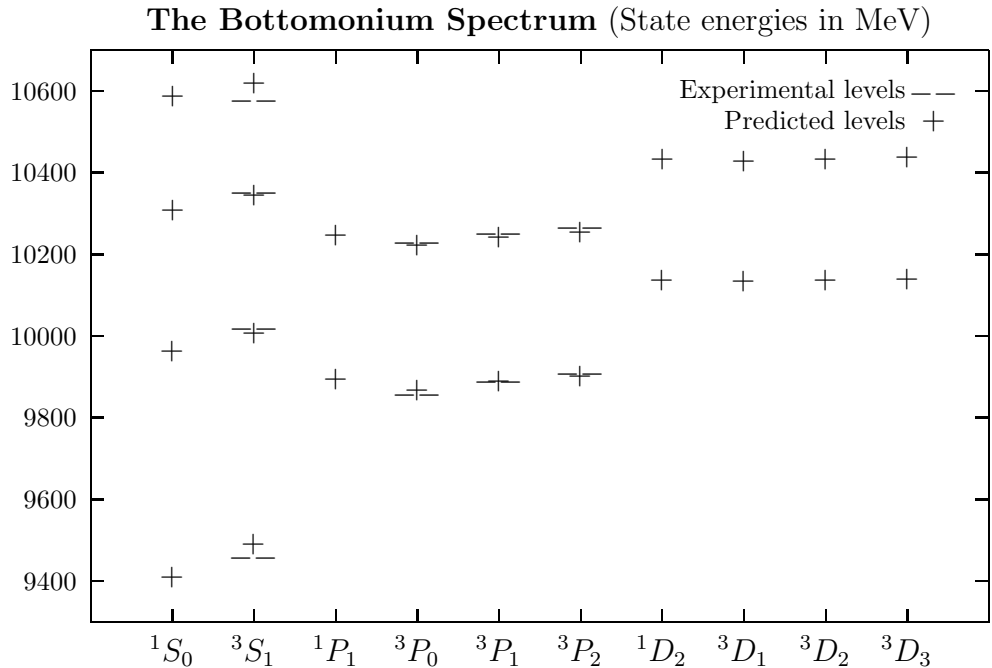
\begin{figure}[ht!]
\begin{center}

\setlength{\unitlength}{0.240900pt}
\ifx\plotpoint\undefined\newsavebox{\plotpoint}\fi
\sbox{\plotpoint}{\rule[-0.200pt]{0.400pt}{0.400pt}}%
\begin{picture}(1650,1034)(0,0)
\font\gnuplot=cmr10 at 10pt
\gnuplot
\sbox{\plotpoint}{\rule[-0.200pt]{0.400pt}{0.400pt}}%
\put(176.0,132.0){\rule[-0.200pt]{4.818pt}{0.400pt}}
\put(154,132){\makebox(0,0)[r]{9400}}
\put(1566.0,132.0){\rule[-0.200pt]{4.818pt}{0.400pt}}
\put(176.0,260.0){\rule[-0.200pt]{4.818pt}{0.400pt}}
\put(154,260){\makebox(0,0)[r]{9600}}
\put(1566.0,260.0){\rule[-0.200pt]{4.818pt}{0.400pt}}
\put(176.0,389.0){\rule[-0.200pt]{4.818pt}{0.400pt}}
\put(154,389){\makebox(0,0)[r]{9800}}
\put(1566.0,389.0){\rule[-0.200pt]{4.818pt}{0.400pt}}
\put(176.0,517.0){\rule[-0.200pt]{4.818pt}{0.400pt}}
\put(154,517){\makebox(0,0)[r]{10000}}
\put(1566.0,517.0){\rule[-0.200pt]{4.818pt}{0.400pt}}
\put(176.0,645.0){\rule[-0.200pt]{4.818pt}{0.400pt}}
\put(154,645){\makebox(0,0)[r]{10200}}
\put(1566.0,645.0){\rule[-0.200pt]{4.818pt}{0.400pt}}
\put(176.0,774.0){\rule[-0.200pt]{4.818pt}{0.400pt}}
\put(154,774){\makebox(0,0)[r]{10400}}
\put(1566.0,774.0){\rule[-0.200pt]{4.818pt}{0.400pt}}
\put(176.0,902.0){\rule[-0.200pt]{4.818pt}{0.400pt}}
\put(154,902){\makebox(0,0)[r]{10600}}
\put(1566.0,902.0){\rule[-0.200pt]{4.818pt}{0.400pt}}
\put(304.0,68.0){\rule[-0.200pt]{0.400pt}{4.818pt}}
\put(304,23){\makebox(0,0){$ ^{1}S_{0}$}}
\put(304.0,946.0){\rule[-0.200pt]{0.400pt}{4.818pt}}
\put(432.0,68.0){\rule[-0.200pt]{0.400pt}{4.818pt}}
\put(432,23){\makebox(0,0){$ ^{3}S_{1}$}}
\put(432.0,946.0){\rule[-0.200pt]{0.400pt}{4.818pt}}
\put(561.0,68.0){\rule[-0.200pt]{0.400pt}{4.818pt}}
\put(561,23){\makebox(0,0){$ ^{1}P_{1}$}}
\put(561.0,946.0){\rule[-0.200pt]{0.400pt}{4.818pt}}
\put(689.0,68.0){\rule[-0.200pt]{0.400pt}{4.818pt}}
\put(689,23){\makebox(0,0){$ ^{3}P_{0}$}}
\put(689.0,946.0){\rule[-0.200pt]{0.400pt}{4.818pt}}
\put(817.0,68.0){\rule[-0.200pt]{0.400pt}{4.818pt}}
\put(817,23){\makebox(0,0){$ ^{3}P_{1}$}}
\put(817.0,946.0){\rule[-0.200pt]{0.400pt}{4.818pt}}
\put(945.0,68.0){\rule[-0.200pt]{0.400pt}{4.818pt}}
\put(945,23){\makebox(0,0){$ ^{3}P_{2}$}}
\put(945.0,946.0){\rule[-0.200pt]{0.400pt}{4.818pt}}
\put(1073.0,68.0){\rule[-0.200pt]{0.400pt}{4.818pt}}
\put(1073,23){\makebox(0,0){$ ^{1}D_{2}$}}
\put(1073.0,946.0){\rule[-0.200pt]{0.400pt}{4.818pt}}
\put(1201.0,68.0){\rule[-0.200pt]{0.400pt}{4.818pt}}
\put(1201,23){\makebox(0,0){$ ^{3}D_{1}$}}
\put(1201.0,946.0){\rule[-0.200pt]{0.400pt}{4.818pt}}
\put(1330.0,68.0){\rule[-0.200pt]{0.400pt}{4.818pt}}
\put(1330,23){\makebox(0,0){$ ^{3}D_{2}$}}
\put(1330.0,946.0){\rule[-0.200pt]{0.400pt}{4.818pt}}
\put(1458.0,68.0){\rule[-0.200pt]{0.400pt}{4.818pt}}
\put(1458,23){\makebox(0,0){$ ^{3}D_{3}$}}
\put(1458.0,946.0){\rule[-0.200pt]{0.400pt}{4.818pt}}
\put(176.0,68.0){\rule[-0.200pt]{339.669pt}{0.400pt}}
\put(1586.0,68.0){\rule[-0.200pt]{0.400pt}{216.328pt}}
\put(176.0,966.0){\rule[-0.200pt]{339.669pt}{0.400pt}}
\put(881,1011){\makebox(0,0){\bf{The Bottomonium Spectrum}\textmd{ (State
energies in MeV)}}}
\put(176.0,68.0){\rule[-0.200pt]{0.400pt}{216.328pt}}
\put(1456,901){\makebox(0,0)[r]{Experimental levels}}
\put(1500,901){\raisebox{-.8pt}{\makebox(0,0){$--$}}}
\put(432,171){\raisebox{-.8pt}{\makebox(0,0){$--$}}}
\put(433,532){\raisebox{-.8pt}{\makebox(0,0){$--$}}}
\put(433,745){\raisebox{-.8pt}{\makebox(0,0){$--$}}}
\put(433,889){\raisebox{-.8pt}{\makebox(0,0){$--$}}}
\put(689,427){\raisebox{-.8pt}{\makebox(0,0){$--$}}}
\put(689,666){\raisebox{-.8pt}{\makebox(0,0){$--$}}}
\put(817,448){\raisebox{-.8pt}{\makebox(0,0){$--$}}}
\put(817,681){\raisebox{-.8pt}{\makebox(0,0){$--$}}}
\put(945,461){\raisebox{-.8pt}{\makebox(0,0){$--$}}}
\put(945,690){\raisebox{-.8pt}{\makebox(0,0){$--$}}}
\put(1456,856){\makebox(0,0)[r]{Predicted levels}}
\put(1500,856){\makebox(0,0){$+$}}
\put(304,139){\makebox(0,0){$+$}}
\put(304,494){\makebox(0,0){$+$}}
\put(305,716){\makebox(0,0){$+$}}
\put(305,895){\makebox(0,0){$+$}}
\put(432,190){\makebox(0,0){$+$}}
\put(433,522){\makebox(0,0){$+$}}
\put(433,739){\makebox(0,0){$+$}}
\put(433,915){\makebox(0,0){$+$}}
\put(561,450){\makebox(0,0){$+$}}
\put(561,676){\makebox(0,0){$+$}}
\put(689,433){\makebox(0,0){$+$}}
\put(689,661){\makebox(0,0){$+$}}
\put(817,446){\makebox(0,0){$+$}}
\put(817,673){\makebox(0,0){$+$}}
\put(945,455){\makebox(0,0){$+$}}
\put(945,681){\makebox(0,0){$+$}}
\put(1073,606){\makebox(0,0){$+$}}
\put(1074,796){\makebox(0,0){$+$}}
\put(1201,604){\makebox(0,0){$+$}}
\put(1202,793){\makebox(0,0){$+$}}
\put(1330,606){\makebox(0,0){$+$}}
\put(1330,796){\makebox(0,0){$+$}}
\put(1458,607){\makebox(0,0){$+$}}
\put(1458,798){\makebox(0,0){$+$}}
\end{picture}

\end{center}
\caption{Bottomonium spectrum with $m_{b}=4720$ MeV
obtained from the Hamiltonian (1)} \label{bb}
\end{figure}

The calculated spectra for the $c\bar c$ and $b\bar b$ systems
that are
obtained by solving the Schr\"odinger equation for the model
Hamiltonian (\ref{ham}) are 
shown in Figs. \ref{cc} and \ref{bb} respectively.
In the
calculation the term of order $p^4$ in the kinetic energy operator
(\ref{kinerg}), as well as 
all the terms of order $m^{-2}$ in the interaction
potentials $V_c$ (\ref{conf}) and $V_G$ (\ref{exch})
were treated in first order 
perturbation theory. The parameter 
values used were those in Table \ref{para}, which
are close to those employed in ref. \cite{Quigg1}. The perturbative
treatment is motivated by the fact that the main focus here will be
the calculation of the radiative widths of the $S$-states, 
for which the only
terms in the fine structure part of the interaction that matter 
are the
delta function terms in (\ref{conf}) and (\ref{exch}), which have to be 
treated perturbatively.\\

\newpage

\begin{table}[ht!]
\begin{center}
\begin{tabular}{|r|r|r|r|r|} \hline
state & $\mathbf{s=0}$ & & $\mathbf{s=1}$ & \\ 
	&	$j=l$	& $j=l-1$ & $j=l$ & $j=l+1$  \\ \hline
\bf{1S}	&	2978	&	&	3159 & \\
        &      (2979)   &       &      (3097) & \\     
\bf{2S}	&	3558	&	&	3683 & \\
        &      (3594)   &       &      (3686) &\\
\bf{3S}	&	3895	&	&	4004 & \\
\bf{4S}  &	4130	&	&	4231 & \\ \hline
\bf{1P}	&	3547	&	3471	&	3538	&	3567	\\
        &      (3526)   &      (3415)   &      (3511)   &      (3556)  \\
\bf{2P}	&	3903	&	3827	&	3892	&	3926	\\
\hline
\bf{1D}	&	3796	&	3798	&	3802	&	3790	\\
\bf{2D}	&	4079	&	4075	&	4083	&	4077	\\
\hline
\end{tabular}
\caption{The $c \bar{c}$-states in MeV.} 
\end{center}
\end{table}

\newpage

\begin{table}[ht!]
\begin{center}
\begin{tabular}{|r|r|r|r|r|} \hline
state & $\mathbf{s=0}$ & & $\mathbf{s=1}$ & \\ 
	&	$j=l$	& $j=l-1$ & $j=l$ & $j=l+1$  \\ \hline
\bf{1S}	&	9411	&	&	9490 & \\
        &               &       &       (9460) &\\
\bf{2S}	&	9964	&	&	10008 & \\
        &               &       &      (10023) &\\
\bf{3S}	&	10311	&	&	10346 & \\
        &               &       &      (10345) &\\
\bf{4S}  &	10589	&	&	10621 & \\ \hline
\bf{1P}	&	9895	&	9868	&	9890	&	9903	\\
        &               &   (9860)    &     (9892)  & (9913) \\
\bf{2P}	&	10248	& 10224	& 10243	& 10256	\\ 
        &               & (10232)  &(10255) & (10268)\\  
\hline
\bf{1D}	&	10138	&	10135	&	10138	&	10140	\\
\bf{2D}	&	10435	&	10431	&	10435	& 10438	\\ \hline
\end{tabular}
\caption{The $b \bar{b}$-states in MeV.} 
\end{center}
\end{table}

The quality of the agreement between the calculated and empirical
$c\bar c$ and $b\bar b$ spectra in Figs. \ref{cc} and \ref{bb} is similar to
that of
the corresponding spectra obtained by the full numerically constructed
Hamiltonian model \cite{Bali1,Bali2}. In both cases the splitting
between the lowest $S$-states ($\eta_c,\, J/\psi)$ is overpredicted
and that of the $P$-states
$(\chi_{cJ})$ is somewhat underpredicted. 
A better description of the $\eta_c-J/\psi$ splitting
may be obtained with the Buchm\"{u}ller-Tye potential
\cite{Buch} used in ref. \cite{Quigg}, but only at the
price of underprediction of the excited states.
This problem appears to be
generic to Hamiltonian models of the 
type (\ref{ham}). The calculated spectra
are nevertheless in good
overall agreement with the empirical spectra, with 
hyperfine splittings of the multiplets which are ordered as
the empirical ones. The model therefore appears realistic enough for a
quantitative calculation 
of the spectrum of the $B_c^+$ system as well \cite{Quigg1}.\\

In Fig. \ref{cc} the empirical $c\bar c$ state
at 4160 MeV has been given the assignment $4 ^3S_1$ state, although
it may be a $2D$ state \cite{Barnes}.
Similarly the empirical $\Upsilon$
state at 10580 MeV has been given the assignment
$4^3 S_1$ in Fig. \ref{bb}. These states fall sufficiently
low for this assignment with the present model, mainly
as a consequence of the $p^2$ term in the effective
confining interaction (\ref{conf}).

\newpage

\subsection{The $B_c^+$ Meson Spectrum} \label{bc-sec}

The first state of the $B_c^+$ meson spectrum, formed of $c$ and $\bar
b$ quarks - presumably a $1^3S_1$-state - has recently been
observed \cite{Fermi}. The Hamiltonian model (\ref{ham}) 
may be applied
directly to the calculation of the spectrum of the $B_c^+$ system. By
setting $\alpha_{s}=0.32$ the $1S$-states agree with the
earlier 
spectra calculated in ref.\cite{Quigg}. 
The precise value for the empirical mass of the discovered
$B_{c}^{+}$-state is not yet known.\\

\begin{figure}[hb!]
\caption{$B_{c}^{+}$ spectrum with $m_{c}=1480$ MeV and $m_{b}=4720$ MeV
obtained from the Hamiltonian (1)}
\label{bc}
\begin{center}

\setlength{\unitlength}{0.240900pt}
\ifx\plotpoint\undefined\newsavebox{\plotpoint}\fi
\sbox{\plotpoint}{\rule[-0.200pt]{0.400pt}{0.400pt}}%
\begin{picture}(1650,500)(0,470)
\font\gnuplot=cmr10 at 10pt
\gnuplot
\sbox{\plotpoint}{\rule[-0.200pt]{0.400pt}{0.400pt}}%
\put(176.0,68.0){\rule[-0.200pt]{4.818pt}{0.400pt}}
\put(154,68){\makebox(0,0)[r]{6200}}
\put(1566.0,68.0){\rule[-0.200pt]{4.818pt}{0.400pt}}
\put(176.0,206.0){\rule[-0.200pt]{4.818pt}{0.400pt}}
\put(154,206){\makebox(0,0)[r]{6400}}
\put(1566.0,206.0){\rule[-0.200pt]{4.818pt}{0.400pt}}
\put(176.0,344.0){\rule[-0.200pt]{4.818pt}{0.400pt}}
\put(154,344){\makebox(0,0)[r]{6600}}
\put(1566.0,344.0){\rule[-0.200pt]{4.818pt}{0.400pt}}
\put(176.0,482.0){\rule[-0.200pt]{4.818pt}{0.400pt}}
\put(154,482){\makebox(0,0)[r]{6800}}
\put(1566.0,482.0){\rule[-0.200pt]{4.818pt}{0.400pt}}
\put(176.0,621.0){\rule[-0.200pt]{4.818pt}{0.400pt}}
\put(154,621){\makebox(0,0)[r]{7000}}
\put(1566.0,621.0){\rule[-0.200pt]{4.818pt}{0.400pt}}
\put(176.0,759.0){\rule[-0.200pt]{4.818pt}{0.400pt}}
\put(154,759){\makebox(0,0)[r]{7200}}
\put(1566.0,759.0){\rule[-0.200pt]{4.818pt}{0.400pt}}
\put(176.0,897.0){\rule[-0.200pt]{4.818pt}{0.400pt}}
\put(154,897){\makebox(0,0)[r]{7400}}
\put(1566.0,897.0){\rule[-0.200pt]{4.818pt}{0.400pt}}
\put(304.0,68.0){\rule[-0.200pt]{0.400pt}{4.818pt}}
\put(304,23){\makebox(0,0){$ ^{1}S_{0}$}}
\put(304.0,946.0){\rule[-0.200pt]{0.400pt}{4.818pt}}
\put(432.0,68.0){\rule[-0.200pt]{0.400pt}{4.818pt}}
\put(432,23){\makebox(0,0){$ ^{3}S_{1}$}}
\put(432.0,946.0){\rule[-0.200pt]{0.400pt}{4.818pt}}
\put(561.0,68.0){\rule[-0.200pt]{0.400pt}{4.818pt}}
\put(561,23){\makebox(0,0){$ ^{1}P_{1}$}}
\put(561.0,946.0){\rule[-0.200pt]{0.400pt}{4.818pt}}
\put(689.0,68.0){\rule[-0.200pt]{0.400pt}{4.818pt}}
\put(689,23){\makebox(0,0){$ ^{3}P_{0}$}}
\put(689.0,946.0){\rule[-0.200pt]{0.400pt}{4.818pt}}
\put(817.0,68.0){\rule[-0.200pt]{0.400pt}{4.818pt}}
\put(817,23){\makebox(0,0){$ ^{3}P_{1}$}}
\put(817.0,946.0){\rule[-0.200pt]{0.400pt}{4.818pt}}
\put(945.0,68.0){\rule[-0.200pt]{0.400pt}{4.818pt}}
\put(945,23){\makebox(0,0){$ ^{3}P_{2}$}}
\put(945.0,946.0){\rule[-0.200pt]{0.400pt}{4.818pt}}
\put(1073.0,68.0){\rule[-0.200pt]{0.400pt}{4.818pt}}
\put(1073,23){\makebox(0,0){$ ^{1}D_{2}$}}
\put(1073.0,946.0){\rule[-0.200pt]{0.400pt}{4.818pt}}
\put(1201.0,68.0){\rule[-0.200pt]{0.400pt}{4.818pt}}
\put(1201,23){\makebox(0,0){$ ^{3}D_{1}$}}
\put(1201.0,946.0){\rule[-0.200pt]{0.400pt}{4.818pt}}
\put(1330.0,68.0){\rule[-0.200pt]{0.400pt}{4.818pt}}
\put(1330,23){\makebox(0,0){$ ^{3}D_{2}$}}
\put(1330.0,946.0){\rule[-0.200pt]{0.400pt}{4.818pt}}
\put(1458.0,68.0){\rule[-0.200pt]{0.400pt}{4.818pt}}
\put(1458,23){\makebox(0,0){$ ^{3}D_{3}$}}
\put(1458.0,946.0){\rule[-0.200pt]{0.400pt}{4.818pt}}
\put(176.0,68.0){\rule[-0.200pt]{339.669pt}{0.400pt}}
\put(1586.0,68.0){\rule[-0.200pt]{0.400pt}{216.328pt}}
\put(176.0,966.0){\rule[-0.200pt]{339.669pt}{0.400pt}}
\put(881,1011){\makebox(0,0){\bf{The $B_{c}+$ Spectrum}\textmd{ (State
energies in MeV)}}}
\put(176.0,68.0){\rule[-0.200pt]{0.400pt}{216.328pt}}
\put(1456,901){\makebox(0,0)[r]{Predicted levels}}
\put(1500,901){\raisebox{-.8pt}{\makebox(0,0){$+$}}}
\put(304,158){\raisebox{-.8pt}{\makebox(0,0){$+$}}}
\put(304,517){\raisebox{-.8pt}{\makebox(0,0){$+$}}}
\put(305,739){\raisebox{-.8pt}{\makebox(0,0){$+$}}}
\put(305,902){\raisebox{-.8pt}{\makebox(0,0){$+$}}}
\put(432,209){\raisebox{-.8pt}{\makebox(0,0){$+$}}}
\put(433,552){\raisebox{-.8pt}{\makebox(0,0){$+$}}}
\put(433,768){\raisebox{-.8pt}{\makebox(0,0){$+$}}}
\put(433,930){\raisebox{-.8pt}{\makebox(0,0){$+$}}}
\put(561,463){\raisebox{-.8pt}{\makebox(0,0){$+$}}}
\put(561,701){\raisebox{-.8pt}{\makebox(0,0){$+$}}}
\put(689,443){\raisebox{-.8pt}{\makebox(0,0){$+$}}}
\put(689,680){\raisebox{-.8pt}{\makebox(0,0){$+$}}}
\put(817,461){\raisebox{-.8pt}{\makebox(0,0){$+$}}}
\put(817,697){\raisebox{-.8pt}{\makebox(0,0){$+$}}}
\put(945,468){\raisebox{-.8pt}{\makebox(0,0){$+$}}}
\put(945,707){\raisebox{-.8pt}{\makebox(0,0){$+$}}}
\put(1073,624){\raisebox{-.8pt}{\makebox(0,0){$+$}}}
\put(1074,817){\raisebox{-.8pt}{\makebox(0,0){$+$}}}
\put(1201,630){\raisebox{-.8pt}{\makebox(0,0){$+$}}}
\put(1202,820){\raisebox{-.8pt}{\makebox(0,0){$+$}}}
\put(1330,628){\raisebox{-.8pt}{\makebox(0,0){$+$}}}
\put(1330,820){\raisebox{-.8pt}{\makebox(0,0){$+$}}}
\put(1458,619){\raisebox{-.8pt}{\makebox(0,0){$+$}}}
\put(1458,814){\raisebox{-.8pt}{\makebox(0,0){$+$}}}
\end{picture}

\end{center}
\end{figure}
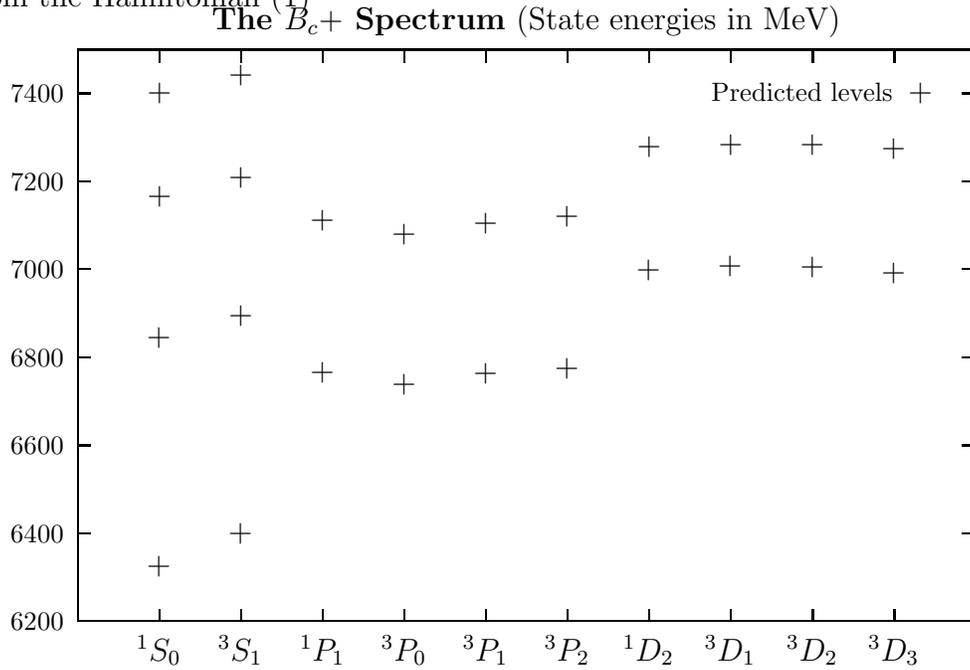

The predicted $B_c^+$ meson spectrum is shown in Fig. \ref{bc}. 
This overall feature of this spectrum
is that of an interpolation between the spectra of the
$c\bar c$ and the $b\bar b$ systems. The large mass difference between
the $c$ and $b$ quarks brings no qualitatively new
features, although it does lead to the antisymmetric spin-orbit
interactions in (\ref{conf}) and (\ref{exch}).\\

\newpage

\begin{table}[ht!]
\begin{center}
\begin{tabular}{|r|r|r|r|r|} \hline
state & $\mathbf{s=0}$ &  & $\mathbf{s=1}$ &  \\ 
	&	$j=l$	& $j=l-1$ & $j=l$ & $j=l+1$  \\ \hline
\bf{1S}	&	6330	&	&	6404 & \\
\bf{2S}	&	6850	&	&	6900 & \\
\bf{3S}	&	7171	&	&	7214 & \\
\bf{4S}  &	7408	&	&	7448 & \\ \hline
\bf{1P}	&	6772	&	6743	&	6769	&	6779	\\
\bf{2P}	&	7116	& 7086	&	7111	&	7125	\\ \hline
\bf{1D}	&	7005	&	7013	&	7010	&	6998	\\
\bf{2D}	&	7284	&	7288	&	7288	&	7280	\\
\hline
\end{tabular}
\caption{The $B_{c}^{+}$-states in MeV.}  \label{BC}
\end{center}
\end{table}

\newpage

The predicted energy values in Table \ref{BC} and in 
Fig. \ref{bc} 
agree fairly well with those obtained in ref.\cite{Quigg1},
although the major shell spacings here are somewhat smaller,
a feature that was built in because that is in better
agreement with the empirical $c\bar c$ and $b\bar b$ spectra.

\section{The Current Operator} \label{curr-sec}

\subsection{Single Quark Current}

Under the assumption that the $b$ and $c$ quarks are point Dirac
particles, their current density operators have the form 

\begin{equation}
<p'|\vec{\jmath}(0)|p>=iQe\bar u(p')\vec \gamma u(p), \label{curr1}
\end{equation}
where $Q$ is the quark charge factor ($+2/3,\,\,-1/3$ respectively).
Assuming canonical boosts for the spinors $u,\bar u$, 
the spin
part of the current (\ref{curr1}) yields the following magnetic moment
operator \cite{Dann}:

\begin{equation}
\vec \mu=Q\left(\frac{m_p}{m_Q}\right)\frac{\vec{\sigma}_Q}
{\sqrt{1+\vec
v^2}}\left\{1-\frac{1}{3}\left(1-\frac{1}{\sqrt{1+\vec
v^2}}\right)\right\}\mu_N. \label{curr2}
\end{equation}
Here $\vec v$ is the quark velocity $\vec v=(\vec p\,'+\vec p)/2m_Q$,
$m_p$ is the proton mass and $\mu_N$ is the nuclear magneton.\\

For the $c\bar c$ and $b\bar b$ systems the spin-magnetic moments
(\ref{curr2}) combine to a spin-flip term of the form 

\begin{equation}
\vec{\mu}_1=\left(\frac{2}{3},-\frac{1}{3}\right)\left(\frac{m_p}
{m_Q}\right)\frac{\vec{
\sigma}_Q-\vec {\sigma}_{\bar Q}}
{\sqrt{1+\vec v^2} }
\left\{1-\frac{1}{3}\left(1-\frac{1}
{\sqrt{1+\vec v^2}}\right)\right\}\mu_N. \label{curr3}
\end{equation}
The matrix element of this operator for a spin-flip transition
$J=1\leftrightarrow J=0$ between $S$-states of the $c\bar c$ and
$b\bar b$ system may be expressed as 
\begin{eqnarray} &&
<f|\vec{\mu}_1|i>=4\left(\frac{2}{3},
-\frac{1}{3}\right)\left(\frac{m_p}
{m_Q}\right)\mu_N\int_{0}^{\infty}dpp^2\int_{-1}^{1}dz
\cr\cr &&
\int_{0}^{\infty}dr'r^{'2}\int_{0}^{\infty}drr^2\varphi_f(r')\frac{1}
{\sqrt{1+\vec v^2}}\left\{1-\frac{1}{3}\left(1-\frac{1}
{\sqrt{1+\vec v^2}}\right)\right
\}\cr\cr &&
j_0\left(r'\left|\vec p+\frac{\vec q}{4}\right|\right)
j_0\left(r\left|\vec p-\frac{\vec q}{4}\right|\right)\varphi_i(r).
\label{curr4}
\end{eqnarray}
Here $\vec v\equiv \vec p/m$, and $\varphi_i(r)$ and $\varphi_f(r')$
are the radial 
wave functions for the initial and final $S$-states.\\

The matrix element (\ref{curr4}) would be 
considerably simplified by expansion
to lowest order in $\vec v$ of the square roots in the expression
(\ref{curr2}). This would however lead 
to misleading results as the radiative
widths of the $S$-states are very sensitive to the model of the current
operator 
and the wave functions \cite{Grotch1}). For reference we note that if
only the order $(v)^0$ term is kept in (\ref{curr3}) the
expression (\ref{curr4})
reduces to the standard non-relativistic impulse approximation 
result

\begin{equation}
<f|\mu |i>=\left(\frac{2}{3},-\frac{1}{3}\right)\left(\frac{m_p}
{m_Q}\right)\mu_N\int_{0}^{\infty}drr^2\varphi_f(r)j_0\left(\frac{qr}
{2}\right)\varphi_i(r).
\end{equation}
Here $Q=c$, $b$ for $c\bar c$ and $b\bar b$ respectively. The higher
order terms in (\ref{curr4}) lead to numerical values
that differ significantly from this result as shown
below.\\

If the magnetic moment expression (\ref{curr4}) is expanded to
second order in $v^2$, the correction factor to 
the non-relativistic
magnetic moment operator is $1-2v^2/3$. The coefficient $2/3$
in this bracket was obtained incorrectly as $5/6$ in refs.
\cite{Grotch1,Grotch2}, with a consequent overestimate of this
relativistic correction. The factor $2/3$ is obtained as
the sum of two terms, one of which arises from the normalization
of the Dirac spinors ($1/2$) \cite{Friar} 
and another, which arises from the
difference between the initial and final quark momenta ($1/6$).\\

For the $B_c^+$ system the spin dependent part of the magnetic
moment operator (\ref{curr3}) is modified to 
\begin{eqnarray}
&&
\vec{\mu}_1(B_c^+)
=\frac{1}{6}m_p(\vec{\sigma}_c+\vec{\sigma}_{\bar b})
\Bigg\{ \frac{2}{m_c} \frac{1}{\sqrt{1+\vec v_c^2}}
\left[1-\frac{1}{3} \left(1-\frac{1}{\sqrt{1+\vec v_c^2}} 
\right) \right] \cr\cr &&
+\frac{1}{m_b} \frac{1}{\sqrt{1+\vec v_b^2}}
\left[1-\frac{1}{3} \left(1-\frac{1}{\sqrt{1+\vec v_b^2}}
\right)\right] \Bigg\}\mu_N\cr\cr &&
+\frac{1}{6}m_p(\vec{\sigma}_c-\vec{\sigma}_{\bar b})
\Bigg\{ \frac{2}{m_c} \frac{1}{\sqrt{1+\vec v_c^2}}
\left[1-\frac{1}{3} \left(1-\frac{1}{\sqrt{1+\vec v_c^2}} 
\right) \right] \cr\cr &&
-\frac{1}{m_b} \frac{1}{\sqrt{1+\vec v_b^2}}
\left[1-\frac{1}{3} \left(1-\frac{1}{\sqrt{1+\vec v_b^2}}
\right)\right] \Bigg\}\mu_N, \label{BCmagm}
\end{eqnarray}
where $\vec v_c\equiv \vec p/m_c$ and $\vec v_b\equiv \vec p/m_b$
respectively. The term that is symmetric in the spins
gives rise to a magnetic moment for the $^3S_1$ state
of the $B_c^+$ system, whereas the term that is 
antisymmetric in the spins gives rise to spin-flip
transitions.\\

The radiative widths calculated using these single
quark magnetic moments -- the impulse approximation -- 
do not agree with the presently known empirical rates
for M1 transitions. These discrepancies can be 
considerably ameliorated, if not entirely eliminated,
by taking into account the exchange current contributions.

\subsection{Exchange Current Operators}

The scalar confining interaction may excite virtual quark-antiquark
states, which are deexcited by the external electromagnetic field.
This process, which is illustrated by the Feynman diagrams in Fig.
\ref{fey},
generates an exchange current -- or two-quark pair excitation current,
which in the $c\bar c$ and $b\bar b$ system may, to order $m^{-2}$, be
expressed in compact form as

\begin{equation}
\vec \jmath_2(C)=-\frac{cr}{m_Q}\vec \jmath_1, \label{curr7}
\end{equation}
where $\vec \jmath_1$ is the 
corresponding single quark current operator. This
exchange current operator is required to satisfy the
continuity equation once the terms of order $1/m^2$ are
included in the confining interaction (\ref{conf}).
\\

The spin dependent
part of the corresponding exchange magnetic moment operator
is then, to lowest order in $1/m^2$, 
for the $c\bar c$ and $b\bar b$ systems:
\begin{equation}
\vec \mu_2(C)=-\left(\frac{2}{3},\, -\frac{1}{3}\right)\frac{m_p}
{m_Q}\frac{cr}{m_Q}(\vec{\sigma}_Q-\vec{ \sigma}_{\bar
q})\mu_N \label{curr8},
\end{equation}
where the factor $2/3$ in the bracket applies to the $c\bar c$ system
and the factor $-1/3$ to the $b\bar b$ system and $m_Q=m_c$ and $m_b$
respectively. The matrix element of this operator for a transition
$i\rightarrow f+\gamma$ is then \cite{Grotch1}
\begin{equation}
<f|\vec \mu_2(C)|i>=
-\left(\frac{2}{3},\, -\frac{1}{3}\right)\frac{m_p}
{m_{Q}^{2}}\mu_N\int_{0}^{\infty}drr^2\varphi_f(r)crj_0\left(\frac{qr}
{2}\right)\varphi_i(r).\label{excur}
\end{equation}
The contributions of these matrix elements turn out to be essential
for the explanation of the empirical decay widths of the transitions
$J/\psi \rightarrow \eta_c\gamma$ and $\psi(2S)\rightarrow
\eta_c\gamma$.\\

In the case of the $B_c^+$ system the proportionality between the
confinement exchange current and the single quark current operator
(\ref{curr7}) is lost. The complete scalar pair excitation current
operator for the $B_c^+$ system, to order $m^{-2}$, has the expression 
\begin{eqnarray} &&
\vec \jmath_2(C)[B_c^+]=  -e\frac{cr}{6}
\Bigg\{2\frac{\vec{ p}_c+\vec{ p}_{\bar c}\,'}
{m_c^2}+\frac{\vec{ p}_{\bar b}+\vec{ p}_{\bar b}\,'}
{m_b^2}\cr\cr && 
 +\frac{i}{2}\left[\left(\frac{2}{m_c^2}+\frac{1}{m_b^2}\right
)(\vec {\sigma}_c+\vec{\sigma}_{\bar b}) 
 +\left(\frac{2}{m_c^2}-\frac{1}{m_b^2}\right)(\vec{ \sigma}_c
-\vec{\sigma}_b)\right]\times \vec q\Bigg\}.
\end{eqnarray}
The spin dependent part of the corresponding magnetic moment operator
takes the form 
\begin{eqnarray} &&
\vec \mu_2(C)[B_c^+]= -\frac{cr}{6}m_p
\cr\cr &&
\left\{\left(\frac{2}
{m_c^2}+\frac{1}{m_b^2}\right)(\vec \sigma_c+\vec \sigma_b) 
 +\left(\frac{2}{m_c^2}-\frac{1}{m_b^2}\right)(\vec \sigma_c-\vec
\sigma_b)\right\}\mu_N. \label{muconf}
\end{eqnarray}
The first term in this expression contributes to the magnetic moment
of the $J>0$ $B_c^+$ mesons.\\

In the $B_c^+$ system the gluon exchange current operator that arises
from excitations of virtual $q\bar q$ pair states will contribute to
both the magnetic moments and to the spin-flip transition strengths.
The complete expression (to order $m^{-2}$) of this gluon exchange
current operator is (in momentum space):
\begin{eqnarray} &&
\vec \jmath_2(G)[B_c^+]=  e\frac{8\pi\alpha_s}{9}
\Bigg\{\frac{2}{3}\frac{1}{k_{\bar
b}^2}\left[\frac{\vec p_{\bar b}+\vec p_{\bar b}\,'} 
{m_cm_b}+i\left(\frac{\vec \sigma_c}{m_c^2}+\frac{\vec \sigma_b}
{m_cm_b}\right)\times\vec k_{\bar b}\right] 
\cr\cr &&
+\frac{1}{3k_c^2}\left[\frac{\vec p_c+\vec p_c\,'}
{m_cm_b}+i\left(\frac{\vec
\sigma_c}{m_cm_b}+\frac{\vec \sigma_b}{m_b^2}\right)\times \vec
k_c\right]\Bigg\}. \label{curr12}
\end{eqnarray}
Here the momentum operators $\vec k_{\bar b}$ and $\vec k_c$
denote the fractional momenta delivered to the $\bar b$ and $c$ quarks
respectively ($\vec q=\vec k_{\bar b}+\vec k_c$).\\

The spin dependent part of the magnetic moment operator for the gluon
exchange current operator (\ref{curr12}) takes the form
\begin{eqnarray} &&
\vec \mu_2(G)[B_c^+]=  \frac{\alpha_s}{27}\frac{m_p}{r}
\Bigg\{\left(\frac{2}{m_c^2}+\frac{3}{m_cm_b}+\frac{1}{m_b^2}\right)(\vec
\sigma_c+\vec \sigma_{\bar b}) \cr\cr &&
+\left(\frac{2}{m_c^2}-\frac{1}{m_cm_b}-\frac{1}{m_b^2}\right)(\vec
\sigma_c-\vec \sigma_{\bar b})\Bigg\}\mu_{N}. \label{curr13}
\end{eqnarray}
Note that the presence of the spin-flip 
term in (\ref{curr13}) is solely a
consequence of the difference in mass of the charm and beauty quarks.\\

\section{Radiative Spin-Flip Transitions} \label{tran-sec}

The spin-flip part of the magnetic moment operators considered above
may be expressed in the general form 
\begin{equation}
\vec \mu_{S.F}={\cal M}(\vec \sigma_Q-\vec \sigma_{\bar
q}),
\end{equation}
where ${\cal M}$ is the matrix element of the orbital part of the
operator for the transition and $e$ is the elementary charge. The
width for a radiative spin-flip transition of the form $Q{\bar
q}(J=1)\rightarrow Q{\bar q}(J=0)\gamma$ then takes the form
\begin{equation}
\Gamma=\frac{4}{3}\alpha_{em}
\frac{M_f}{M_i}{\cal M}^2q^3. \label{sf2}
\end{equation}
Here $\alpha_{em}$ is the fine structure constant, $q$ is the photon
momentum in the laboratory frame, and $M_i$ and $M_f$ are the masses
of the initial and final states respectively. The matrix element
${\cal M}$ is formed of a single quark current term and an exchange
current term. These expressions for the heavy quarkonium
systems are given explicitly below.\\

\subsection{Spin-flip transitions in the $c\bar c$ system}

For radiative spin-flip decays of the form $\psi(nS)\rightarrow
\eta_c\gamma$ 
and $\eta_c'(nS)\rightarrow \psi \gamma$
the matrix element (\ref{sf2}) ${\cal M}$ may be derived from
the single quark operator (\ref{curr3}) 
and the exchange current contribution
(\ref{curr8}) that is associated with the scalar confining interaction. 
The matrix element ${\cal M}$ then takes the form \cite{Grotch1}:
\begin{equation}
{\cal M}=\frac{1}{3m_c}\{I_1+I_c\}.\label{matrcc}
\end{equation}
The  factor $1/3m_c$ arises from the charge factor $2/3$ and
the factor $1/2m_c$ in the magnetic moment operator. The dimensionless 
integrals $I_1$ and $I_c$ that arise from the
single quark operator (\ref{curr3}) and the confining exchange current 
(\ref{curr8})
respectively have the explicit expressions (cf.(\ref{curr4}),
(\ref{excur})).
\begin{eqnarray} &&
I_1=4\int_{0}^{\infty}dpp^2\int_{-1}^{1}dz\int_{0}^{\infty}
dr'r'^2\int_{0}^\infty dr r^2 \varphi^*_f(r')\cr\cr
&&
\frac{1}{\sqrt{1+v^2}}\left\{1-\frac{1}{3}\left(1-\frac{1}
{\sqrt{1+v^2}}\right)\right\}
\cr\cr &&
j_0\left(r'\sqrt{p^2+pqz/2+q^2/16}\right)j_0\left(r\sqrt{p^2-pqz/2+q^2/16}
\right)\varphi_i(r),\label{I1}
\end{eqnarray}
\begin{equation}
I_c=-\frac{c}
{m_c}\int_{0}^{\infty}drr^2\varphi_f^*(r)r\varphi_i(r).
\label{IC}
\end{equation}
Here $v\equiv p/m$. The first of these two integrals reduces to the
form (\ref{curr4}) 
(with exception of the factor $2m_p\mu/3m_Q$) in the static
limit $v\rightarrow 0$. \\

\begin{table}[h!]
\begin{center}
\begin{tabular} {l|r|r|r|c|} 
Transition &NRIA & RIA& RIA+conf & Exp. \\ \hline
$1^{3}S_{1}\rightarrow 1^{1}S_{0}$ & 
2.87 keV & 2.18 keV
&1.05 keV & 1.14 $\pm$ 0.39 keV\\ \hline
$2^{3}S_{1}\rightarrow 1^{1}S_{0}$ 
&1.54 keV & 0.206 keV
& 1.32 keV & 0.78 $\pm$ 0.24 keV\\ 
$2^{3}S_{1}\rightarrow 2^{1}S_{0}$ 
&1.43 keV & 1.03 keV
& 0.14 keV & --- \\ \hline
$2^{1}S_{0}\rightarrow 1^{3}S_{1}$ 
&0.188 keV & 0.469 keV
& 0.351 keV & --- \\ \hline
$3^{3}S_{1}\rightarrow 1^{1}S_{0}$ 
& 0.795 keV & 0.172 keV
& 0.929 keV & --- \\
$3^{3}S_{1}\rightarrow 2^{1}S_{0}$ 
& 0.61 keV & 0.0753 keV 
& 1.24 keV & --- \\
$3^{3}S_{1}\rightarrow 3^{1}S_{0}$ 
& 2.32 keV & 1.57 keV
& 0.0124 keV & --- \\ \hline
$3^{1}S_{0}\rightarrow 1^{3}S_{1}$ 
& 0.167 keV & 0.228 keV
& 0.372 keV & --- \\
$3^{1}S_{0}\rightarrow 2^{3}S_{1}$ 
& 0.0133 keV & 0.12 keV 
& 0.233 keV & --- \\ \hline
\end{tabular}
\caption{Decay widths for the $c \bar{c}$ system. The columns
NRIA and RIA contain the results of the non-relativistic
and relativistic impulse approximations respectively. The
net calculated decay width obtained by combining the
relativistic impulse approximation and exchange current
contributions is given in the column RIA+conf.
The empirical values are from ref.\cite{PDG}}\label{wcc}
\end{center}
\end{table}

The calculated decay widths for spin-flip transitions between
the $S-$states of the $c\bar c$ system are given in 
Table \ref{wcc}. In Table \ref{wcc} the decay width results
that are obtained in the non-relativistic and relativistic
impulse approximations are also given both in order to
facilitate comparison to the results of refs. \cite{Grotch1,
Grotch2} and to bring out the significant role of the
exchange current contribution. Note that in the numerical
calculation the empirical values for the photon momenta
have been used in order to obtain the correct phase
space factors. This makes a difference only for transitions
that involve the $J/\psi$ state, which is somewhat
overpredicted by the present model. This issue does not
affect the wave functions, which were calculated without
account of the hyperfine terms of order $v^2/c^2$, which
were treated in lowest order perturbation theory.\\

The results in Table \ref{wcc} show that the non-relativistic
impulse approximation overestimates the rate for
$J/\psi\rightarrow \eta_c\gamma$ by a factor $\sim 3$ as has
been noted earlier \cite{Grotch1,Grotch2}. By employing
the unapproximated magnetic moment operator (\ref{curr3})
this overestimate is reduced to a factor $\sim 2$. Agreement
with the empirical value is achieved only by taking into
account the exchange current contribution. In the case of
the transition $\psi(2S)\rightarrow \eta_c\gamma$
the impulse approximation underestimates the
empirical decay width by a large factor. For this 
transition the exchange current contribution is the
dominant one. The net calculated result falls only
slightly above the upper limit of the uncertainty
range of the present empirical result. There is no
instance, where the non-relativistic impulse approximation
result is close to either the empirical value, or
to the net calculated result. This feature 
emphasizes the point that the M1 transitions are
peculiarly sensitive to the strength and form of
the confining interaction through the associated
exchange current. The ultimate reason for this
sensitivity is of course the fact that there is
destructive interference between the different
contibutions to the transition amplitude. Those
calculated widths, which are exceptionally small
therefore also have the largest theoretical
uncertainty, as even tiny modifications in the
model parameters could cause large relative changes.\\ 

\subsection{Spin-flip transitions in the $b\bar{b}$ system}

For radiative spin-flip transitions in the 
$b\bar b$ system the matrix element ${\cal M}$ in (\ref{sf2}) 
takes the form :
\begin{equation}
{\cal M}=-\frac{1}{6m_b}\{I_1+I_c\}. \label{matrbb}
\end{equation}
The  factor $-1/6m_b$ arises from the charge factor $-1/3$ and
the factor $1/2m_b$ in the magnetic moment operator. The dimensionless 
integrals $I_1$ and $I_c$ are defined as in eqs. (\ref{I1}),
(\ref{IC}).\\

The calculated decay rates reveal that even in the case
of the $b\bar b$ system the non-relativistic impulse
approximation is unreliable, except in the case, where the
initial and final states have the same degree of excitation.
In those cases it leads to overestimates of the net
calculated decay rates by factors $\sim 2$. The exchange
current contribution is large for all these spin-flip
transitions, and in fact the dominant contribution
for transitions between states with different excitation
number.\\

\begin{table}[h!]
\begin{center}
\begin{tabular} {l|r|r|r|}
Transition & NRIA & RIA & RIA + conf \\ \hline 
$1^{3}S_{1}\rightarrow 1^{1}S_{0}$ 
& 23.4 eV & 21.2 eV
& 18.9 eV \\ \hline
$2^{3}S_{1}\rightarrow 1^{1}S_{0}$ 
& 3.45 eV & 2.02 eV
& 0.215 eV \\
$2^{3}S_{1}\rightarrow 2^{1}S_{0}$ 
& 4.09 eV & 3.70 eV
& 2.77 eV \\ \hline 
$2^{1}S_{0}\rightarrow 1^{3}S_{1}$ 
& 0.74 eV & 2.25 eV
& 0.0069 eV \\ \hline
$3^{3}S_{1}\rightarrow 1^{1}S_{0}$ 
& 2.82 eV & 3.56 eV
& 0.126 eV \\
$3^{3}S_{1}\rightarrow 2^{1}S_{0}$ 
& 1.41 eV & 0.35 eV
& 1.09 eV \\
$3^{3}S_{1}\rightarrow 3^{1}S_{0}$ 
& 2.06 eV & 1.83 eV
& 1.32 eV \\ \hline
$3^{1}S_{0}\rightarrow 1^{3}S_{1}$ 
& 1.15 eV & 3.50 eV
& 0.347 eV \\
$3^{1}S_{0}\rightarrow 2^{3}S_{1}$ 
& 0.296 eV & 0.529 eV
& 0.253 eV \\ \hline
\end{tabular}
\caption{Decay widths for the $b \bar{b}$ system. The columns
NRIA and RIA contain the results of the non-relativistic
and relativistic impulse approximations respectively. The
net calculated decay width obtained by combining the
relativistic impulse approximation and exchange current
contributions is given in the column RIA+conf.}
\end{center}
\end{table}

\subsection{Spin-flip transitions in the $B_{c}^{+}$ system}

The magnetic moment of the $^3S_1$ ground state of the
$B_C^+$ system that is obtained from the expression
\ref{BCmagm} is
\begin{eqnarray} &&
\mu(RIA)=\frac{4}{3}m_p
\mu_N\int_{0}^{\infty}dpp^2\int_{-1}^{1}dz
\cr\cr &&
\int_{0}^{\infty}dr'r^{'2}\int_{0}^{\infty}drr^2\varphi(r')
\Bigg\{\frac{2}{m_c
\sqrt{1+\vec v_c^2}}\left\{1-\frac{1}{3}\left(1-\frac{1}
{\sqrt{1+\vec v_c^2}}\right)\right\}
\Bigg\}\cr\cr &&
+\Bigg\{\frac{1}{m_b
\sqrt{1+\vec v_b^2}}\left\{1-\frac{1}{3}\left(1-\frac{1}
{\sqrt{1+\vec v_b^2}}\right)\right\}\Bigg\}\cr\cr &&
j_0(r'p)
j_0(r p)\varphi(r).
\label{bcmag}
\end{eqnarray}
Here $\varphi(r)$ is the radial wave function for the
$^3S_1$ state. If the velocity dependent ``relativistic''
correction terms are dropped from this expression 
it reduces to the static quark model result
\begin{equation}
\mu(IA)\simeq \frac{1}{3}\bigg\{\frac {2 m_p}{m_c}+
\frac{m_p}{m_b}\bigg\}\mu_N.\label{static}
\end{equation} 
With the quark mass values in Table \ref{para} the
static approximation value is 0.49 $\mu_N$. When
calculated from the unapproximated expression (\ref{bcmag})
this result is reduced to 0.420 $\mu_N$.\\

The exchange current contributions to this magnetic
moment value is obtained from the spin-symmetric
terms in the exchange magnetic moment expressions
\ref{muconf} and \ref{curr13} for the confining and
gluon exchange interactions respectively. The explicit
expressions for these corrections are
\begin{equation}
\mu(C)=-\frac{m_p}{3}c
\bigg\{ \frac{2}{m_c^2}+\frac{1}{m_b^2} \bigg\}
\int_0^\infty dr r^3 \varphi^2(r)
\mu_N,
\label{muC}
\end{equation}
\begin{equation}
\mu(G)=\frac{2 \alpha_s}{27} m_p
\bigg\{ \frac{2}{m_c^2}+\frac{3}{m_c m_b}+\frac{1}{m_b^2}
\bigg\}
\int_0^\infty dr r\varphi^2(r) \mu_N.\label{muG}
\end{equation}
Numerical evaluation of these two expressions
give the results $\mu(C)=-0.10 \mu_N$ and 
$\mu(G)=0.036 \mu_N$ respectively. When
these values are added to the impulse approximation
result 0.420 $\mu_N$ the net calculated magnetic moment becomes
0.35 $\mu_N$. The static quark model value therefore
is expected to represent an overestimate of
the net magnetic moment value of about 23 \%.
This conclusion is in line with estimates of the
exchange current and relativistic corrections to the
static quark model predictions of the magnetic
moments of the baryons, but in that case the
situation is more complex because of the
possibly substantial exchange current contributions
associated with flavor dependent meson exchange
interactions \cite{Dann,Helminen}.\\

The spin-flip transitions in the $B_{c}^+$ systems differ from
those of the $c\bar c$ and $b\bar b$ systems in that they
also obtain contributions from the gluon exchange magnetic
moment operator (\ref{curr13}). This contribution turns out in most
cases to be weak in comparison with that from the exchange
magnetic moment operator that is due to the confining
interaction (\ref{muconf}). This is shown in Table \ref{wbc},
where the calculated decay widths for the spin-flip 
transitions between the (predicted) $S-$states of the
$B_c^+$ system are given with and without inclusion of the
gluon exchange current contribution. The role of the gluon
exchange current contribution is in every instance to
increase the net exchange current contribution.\\

In the non-relativistic impulse approximation we obtain
spin-flip transition decay widths that are close
to those obtained in ref.\cite{Quigg1}. As shown by
the comparison between the calculated decay widths
that are obtained in the non-relativistic impulse
approximation with the values obtained
with the complete Dirac magnetic moment operator,
the former approximation leads to overestimates by
factors, which are typically larger than $\sim 2$.\\

The role of the exchange current contribution is
predicted to be
large for the spin-flip transitions in the
$B_c^+$ system. The net calculated result typically
differs from that obtained in the relativistic
impulse approximation
by an order of magnitude, the only exceptions being
the transitions between states with the same
excitation number. \\

\begin{table}[h!]
\begin{center}
\begin{tabular} {l|r|r|r|r|}
Transition & NRIA & RIA & RIA + conf & RIA +conf + gluon \\ \hline
$1^{3}S_{1}\rightarrow 1^{1}S_{0}$ 
& 138 eV & 91.4 eV & 42.5 eV
& 51.4 eV \\ \hline
$2^{3}S_{1}\rightarrow 1^{1}S_{0}$ 
& 62.1 eV & 375 eV & 5.40 eV
& 46.1 eV \\ 
$2^{3}S_{1}\rightarrow 2^{1}S_{0}$ 
& 43.0 eV & 27.0 eV & 3.08 eV
& 3.94 eV \\ \hline
$2^{1}S_{0}\rightarrow 1^{3}S_{1}$ 
& 12.6 eV & 249 eV & 0.688 eV 
& 15.6 eV \\ \hline
$3^{3}S_{1}\rightarrow 1^{1}S_{0}$ 
& 38.5 eV & 145 eV & 30.5 eV
& 129 eV \\
$3^{3}S_{1}\rightarrow 2^{1}S_{0}$ 
& 21.3 eV & 110 eV & 56.9 eV
& 85.7 eV \\
$3^{3}S_{1}\rightarrow 3^{1}S_{0}$ 
& 27.4 eV & 15.8 eV & 0.0145 eV
& 0.0758 eV \\ \hline
$3^{1}S_{0}\rightarrow 1^{3}S_{1}$ 
& 14.3 eV & 107 eV & 17.8 eV
& 80.8 eV \\
$3^{1}S_{0}\rightarrow 2^{3}S_{1}$ 
& 3.00 eV & 66.5 eV & 23.8 eV
& 35.8 eV \\ \hline
\end{tabular}
\caption{Decay widths for the $B_{c}^{+}$ system}\label{wbc}
\end{center}
\end{table}

\section{Discussion} \label{disc-sec}

The present results suggest that the exchange current
operator that is associated with the scalar confining
interaction through current conservation gives a
crucial contribution to the decay rates for radiative
spin-flip transitions  in heavy quarkonia. In most
cases it interferes destructively with the matrix
element of the single quark magnetic moment operator.
As a consequence the value for these decay rates obtained
in the impulse approximation differs from the net calculated
decay rate by large factors.\\

The results support the observation made in refs.
\cite{Grotch1,Grotch2} that the relativistic corrections
to the single quark magnetic moment operator are of
significant magnitude, although with the present
unapproximated treatment of the relativistic
magnetic moment operator, these relativistic corrections
are moderated somewhat.\\

The most important exchange current operator is that
associated with the scalar confining interaction. The
gluon exchange current only contributes a small 
additional correction to the radiative decays of the 
$B_c+$ system. Both these exchange current operators
involve excitation of intermediate virtual
$q\bar q$ pairs. The exchange current corrections therefore
do involve the negative energy components of the
quark spinors. That such negative energy components are
numerically significant has also been pointed out in
ref.\cite{Snellman}, although in a quite different
approach to the problem, based on an instantaneous
approximation to the Bethe-Salpeter equation. The
present Schr\"{o}dinger equation approach with
exchange currents, pioneered in refs.\cite{Grotch1,Grotch2},
would appear somewhat more transparent, and as
shown above appears to yield decay rates for the
spin-flip transitions of heavy quarkonia, that
compare more favorably with extant empirical values.
The present Schr\"{o}dinger equation approach may
be derived from the Bethe-Salpeter equation
by means of a three-dimensional quasipotential
reduction in the adiabatic limit \cite{Coester}. 
\\ 

The present results suggest that the exchange current
operator that is associated with the confining
interaction should play an important role in
radiative transitions of heavy vector mesons to
heavy pseudoscalar mesons, in particular the
transitions 
$D^*\rightarrow D\gamma$, $D_s^*\rightarrow D_s\gamma$
and $B^*\rightarrow B\gamma$.\\

\centerline{\bf Acknowledgments}

\vspace{0.5cm}

DOR thanks Dr. H\aa kan Snellman for drawing attention to
the M1 transitions in heavy quarkonia. TL and CN thank
the Finnish Society of Sciences and Letters for support.
Research supported in part by the Academy of
Finland under contract 34081.\\

\end{document}